\documentclass[%
 superscriptaddress,
 preprint,
amsmath,amssymb,
aps,
prb,
floatfix,
]{revtex4-2}

\usepackage[english]{babel}

\usepackage[dvipsnames]{xcolor}
\usepackage{graphicx}
\usepackage{dcolumn}
\usepackage{bm}
\usepackage{upgreek}

\usepackage{amssymb}
\usepackage{amsmath}
\usepackage{textcomp}
\usepackage{soul}

\usepackage{xr-hyper}
\makeatletter
\newcommand*{\addFileDependency}[1]{
  \typeout{(#1)}
  \@addtofilelist{#1}
  \IfFileExists{#1}{}{\typeout{No file #1.}}
}
\makeatother

\newcommand*{\myexternaldocument}[1]{
    \externaldocument{#1}
    \addFileDependency{#1.tex}
    \addFileDependency{#1.aux}
}

\myexternaldocument{SupInfo}

\usepackage{hyperref}

\begin{document}


\title{Possible evidence for Harper broadening in the yellow exciton series of \texorpdfstring{Cu$_2$O}{Cu2O} at ultrahigh magnetic fields}

\author{Zhuo Yang}
\affiliation{Institute for Solid State Physics, The University of Tokyo, Kashiwa, Chiba, 277-8581, Japan}

\author{Jinbo Wang}
\affiliation{Institute for Solid State Physics, The University of Tokyo, Kashiwa, Chiba, 277-8581, Japan}

\author{Yuto Ishii}
\affiliation{Institute for Solid State Physics, The University of Tokyo, Kashiwa, Chiba, 277-8581, Japan}

\author{Duncan K. Maude}
\affiliation{Laboratoire National des Champs Magn\'etiques Intenses, CNRS-UGA-UPS-INSA, 143 avenue de Rangueil, 31400 Toulouse, France}

\author{Atsuhiko Miyata}
\affiliation{Institute for Solid State Physics, The University of Tokyo, Kashiwa, Chiba, 277-8581, Japan}

\author{Yasuhiro H. Matsuda}
\email{ymatsuda@issp.u-tokyo.ac.jp}
\affiliation{Institute for Solid State Physics, The University of Tokyo, Kashiwa, Chiba, 277-8581, Japan}

\date{\today}

\newcommand{\Joey}[1]{{\color[RGB]{203,5,5}{#1}}}
\newcommand{\Joeyc}[1]{{\color[RGB]{203,5,5}{[Joey: {#1}\,]}}}
\newcommand{\Joeyx}[1]{{\color[RGB]{55,173,107}{\st{#1}}}}

\newcommand{\Yhmc}[1]{{\color[RGB]{184,0,184}{[Matsuda: {#1}\,]}}}
\newcommand{\Yhm}[1]{{\color[RGB]{184,0,184}{#1}}}
\newcommand{\Yhmx}[1]{{\color[RGB]{184,0,184}{\st{#1}}}}

\clearpage

\begin{abstract}
Hydrogen-like systems in ultra-high magnetic fields are of significant interest in interdisciplinary research. Previous studies have focused on the exciton wavefunction shrinkage under magnetic fields down to artificial crystal lattices (e.g., quantum wells, superlattices), where the effective mass approximation remains valid. However, further compression toward the natural crystal lattice scale remains experimentally challenging.
 In this study, we report magneto-absorption measurements on the yellow-exciton series in Cu$_2$O using pulsed magnetic fields of up to 500\,T. The strong low energy absorption features are assigned to the spin Zeeman split 2p$_{0}$ and 3p$_{0}$ exciton states. The high field data provides a value for the reduced effective mass of the exciton $\mu^* = 0.415 \pm 0.01 m_e$. Intriguingly, the broadening of the 2p$_{0}$ ground state transition exhibits a sudden increase for ultrahigh magnetic fields above 300\,T, providing possible evidence for Harper broadening - an indication of the breakdown of the effective mass approximation when the magnetic length becomes comparable to the lattice constant of the crystal.

\end{abstract}

\maketitle

\clearpage

\section{Introduction}

Hydrogen-like systems, consisting of a positive and a negatively charged particle that are bound via the Coulomb interaction, have been a central topic in high magnetic field research for more than 70 years.  This subject is of significant interest in interdisciplinary research areas due to the unique behavior of hydrogen under extreme conditions. In astrophysics, the shrinkage of the hydrogen atom’s wave function in ultrahigh magnetic fields plays a crucial role in understanding the reorganization of chemical bonding between the hydrogen atoms, forming a one dimensional H$_n$ molecular chain along the magnetic field direction\,\cite{ruderman1971matter,chui1974excitonic,salpeter1998hydrogen}. This reorganization determines the atmospheric conditions on white dwarf and neutron stars, where extremely high magnetic fields are naturally present\,\cite{garstang1977atoms}. In condensed matter physics, excitons - analogous hydrogen-like systems in semiconductors - are studied under high magnetic fields to reveal fundamental material properties, such as the exciton binding energy, dielectric screening, and effective mass \,\cite{Miyata2015MAPbI3,Galkowski2016FAPbI3,Yang2017CsPbI3,Yang2017MAPbI3,soufiani2017impact}. A detailed knowledge of these parameters is essential for designing and optimizing advanced optoelectronic devices. 

Numerous theoretical studies have explored the energy evolution and wave function compression of hydrogen-like systems in ultrahigh magnetic fields\,\cite{murdin2013si,zielinska2019magneto}. Early experimental results focus on the investigation of the magnetic field induced exciton wave function shrinkage down to the size of artificial crystal lattices, for example few nanometer wide quantum wells or superlattices\,\cite{tarucha1984exciton,miura1998magneto,miura2005pulsed,albrecht2001evidence}, in which case the effective-mass theory remains valid\,\cite{sasaki1990high}. However, due to the limited strength of laboratory scale magnetic fields, shrinking the exciton wave function down to the natural (real) crystal lattice constant (a few angstroms), remains extremely challenging. Electromagnetic flux compression (EMFC) is a pulsed field generation technique that can produce a magnetic field of up to 1200\,T, sufficiently high to compress the hydrogenic wave function down to the crystal lattice dimensions\,\cite{nakamura2018record}. Crucially, this technique is also very well suited to magneto-optical measurements - the most widely used methods for investigating hydrogen-like systems. 

Cu$_2$O is a direct band gap semiconductor with well-known material properties, \emph{e.g.} a band gap energy of $\simeq 2.17$\,eV, and a reduced effective mass of $\simeq 0.4\,m_e$\,\cite{artyukhin2018magneto,zhilich1969magnetoabsorption,hodby1976cyclotron,halpern1967energy}. The conduction and valence bands originate from the $4s$ and $3d$ orbitals of Cu, respectively\,\cite{elliott1961Symmetry}. The yellow exciton series in Cu$_2$O is archetypical of Wannier-Mott excitons\,\cite{wannier1937structure}, with a relatively large excitonic binding energy of $\simeq 98$\,meV\,\cite{kazimierczuk2014giant,Matsumoto1996Revived}. This makes Cu$_2$O an ideal system to investigate hydrogen-like systems in ultra high magnetic fields.

In this investigation, we have performed spectrally resolved magneto-absorption on a single crystal of Cu$_2$O at low temperature using different field generation techniques, including traditional non-destructive pulsed magnets (NDPM), single-turn coil (STC), and EMFC methods, covering the magnetic field range from 0 to 520\,T. The same single crystal was used for all measurements facilitating a comparison between data taken using the different techniques.The ultra-high field absorption measurement helps identify the quantum numbers of the pronounced 2p absorption peaks as the spin Zeeman split 2p$_{0}$ state, and determine the reduced mass of the  exciton $\mu^* = 0.415 \pm 0.01\,m_e$. Interestingly, the full width half maximum (FWHM) of the 2p$_{0}$ ground state transition exhibit a sudden increase at fields above 300\,T, accompanied by an increase of reduced effective mass of exciton. These features are consistent with the expected Harper broadening, predicted using a tight binding approximation, for electrons moving in the conduction band of a metal in a uniform magnetic field\,\cite{harper1955single}.  This suggests a break-down of the effective mass approximation when the magnetic length becomes comparable to the lattice constant of the crystal.

\section{Experimental results}

\begin{figure*}[t!]
\centering
\includegraphics[width= 0.9\linewidth]{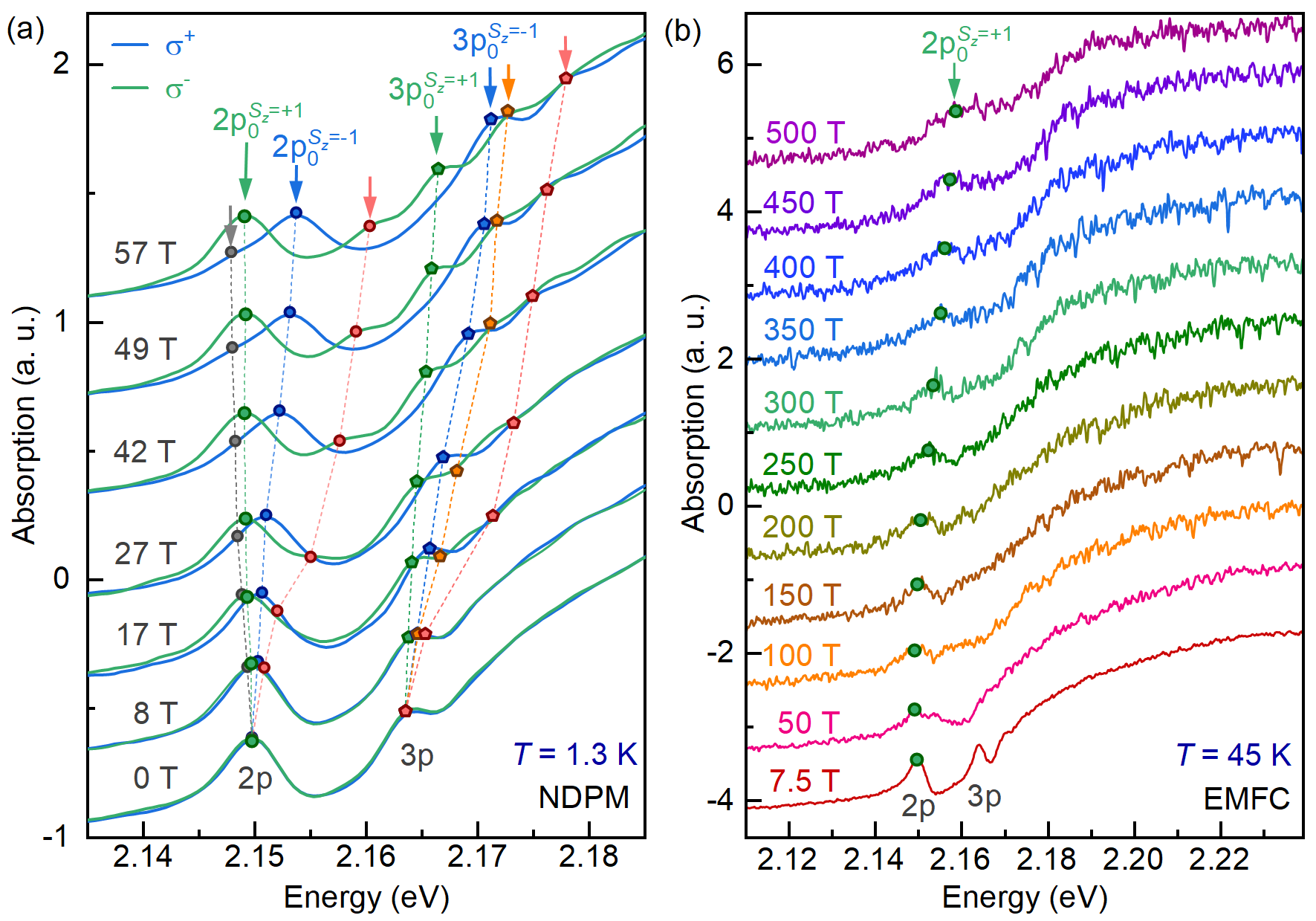}
\caption{(a) Exciton absorption spectra of Cu$_2$O for both $\sigma^+$ (blue lines) and $\sigma^-$
(green lines) measured in NDPM with fields up to 60\,T. (b) Exciton absorption spectra of Cu$_2$O measured in EMFC with field up to 500\,T. Both experiments were performed on the same sample in the Faraady geometry. The spectra are shifted vertically for clarity. 
}
\label{Fig:FieldSpectra}
\end{figure*}

Fig.\,\ref{Fig:FieldSpectra}(a) shows the magneto-absorption spectra of Cu$_2$O measured at 1.3\,K and at magnetic fields of up to 60\,T using a NDPM technique. A broad band circular polarizer was used to identify the $\sigma^+$ and $\sigma^-$ absorption, shown as blue and green solid lines. As the field increases, the 2p and 3p states split into two pronounced peaks (marked by blue and green arrows), corresponding to $\sigma^+$ and $\sigma^-$ polarization, respectively. In addition to the two strong absorption lines, the spectra exhibit several shoulder like features in the magnetic field above 27\,T (marked as gray, orange and red arrows). These shoulder features (peaks) were also resolved by Artyukhin et al.\,\cite{artyukhin2018magneto} by measuring high-resolution magneto-absorption in a DC magnetic field up to 31\,T. A comparison of the absorption energy of these spin-split 2p and 3p states versus magnetic field of Artyukhin \emph{et al.} and from this work is shown in Fig.\,\ref{SI:Fig:DatapointsCom}\,(a-b). The two sets of data agree very well within the experimental error.

\begin{figure*}[t!]
  \centering
   \includegraphics[width= 0.8\linewidth]{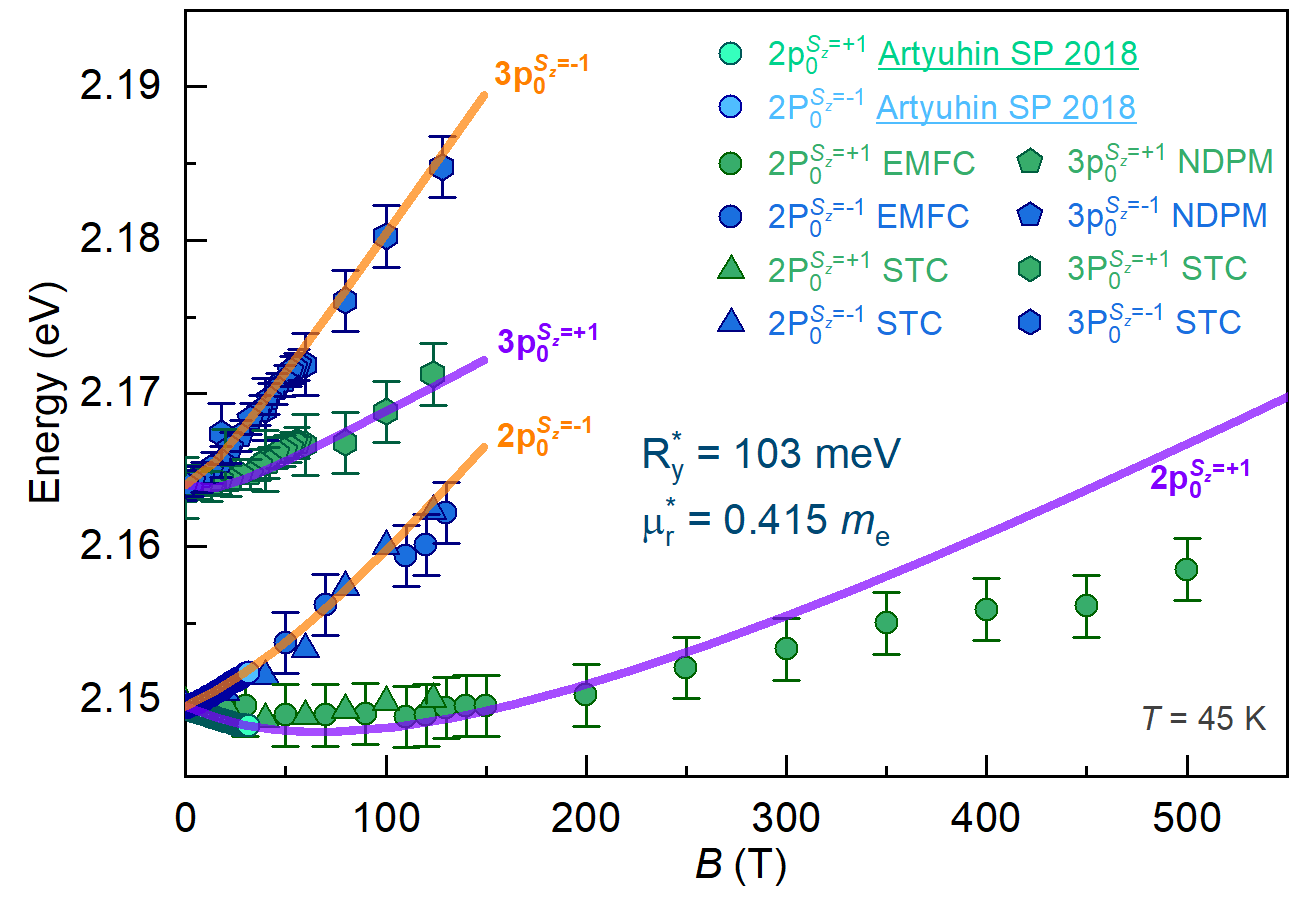}
  \caption{
 Energies for 2p$_{0}^{S_z=\pm1}$ and 3p$_{0}^{S_z=\pm1}$ excitonic absorption as a function of applied magnetic field at 45\,K. The green and blue symbols indicate the absorption energy determined from $\sigma^+$ and $\sigma^-$ polarizations. Data points below 32\,T, taken from ref\,\cite{artyukhin2018magneto}, are shown as light blue and green closed circles.  The orange and violet curves are fits to the excitonic absorption energy using the hydrogen model in a magnetic field. The energy evolution of 2p$_0^{S_z=+1}$  deviates from the hydrogen model at fields above 300\,T
  }
  \label{Fig:energy2p3p}
\end{figure*}

To keep track on the evolution of the excitonic states in higher field region, we performed magneto-absorption measurements on the same sample using EMFC technique with magnetic fields up to 500\,T, as shown in Fig.\,\ref{Fig:FieldSpectra}(b). In such ultra-high magnetic field, only one of the split 2p levels is clearly resolved in the high field spectra, which was later identified as 2p$_0^{S_z=+1}$ level from the spin Zeeman split orthoexciton (marked as green symbol in Fig.\,\ref{Fig:FieldSpectra}(b)). Fig.\,\ref{Fig:energy2p3p} summarizes the evolution of the absorption energies of 2p and 3p excitonic states as a function of magnetic field, measured using different field generation techniques, including data points taken from the exquisitely detailed magneto-absorption measurements of Artyuhin \emph{et al.} in DC magnetic fields\,\cite{artyukhin2018magneto}. The transition energies determined from the different field generation techniques, and from the literature, exhibit a good consistency.

Under magnetic field, the 2p level split into four absorption peaks in the Faraday geometry and are identified by $\sigma^+$ and $\sigma^-$ circular polarization, respectively (Fig.\,\ref{Fig:FieldSpectra}(a)). Here, we focus on the two pronounced absorption peaks that are observable in magnetic fields above 100\,T. Their absorption energy are marked as green and blue circle symbols in Fig.\,\ref{Fig:energy2p3p}. These two absorption peaks are generally identified in the literature as 2p$_{+1}$ and 2p$_{-1}$ levels with magnetic quantum number of $m=\pm 1$\,\cite{kobayashi1989yellow,zielinska2019magneto}. However, based on this assignment and the observed splitting of these two levels in magnetic field, the extracted reduced mass value would be $\simeq 2\,m_\text{e}$ (see SI Sec.\,I), which is five times larger than the accepted literature value\,\cite{artyukhin2018magneto,zhilich1969magnetoabsorption,halpern1967energy,hodby1976cyclotron}. The exciton is intrinsically a coupled two spin system, and the total spin is either $S=1$ (orthoexciton or triplet state) or $S=0$ (paraexciton or singlet state)\,\cite{blundell2001magnetism}. In this study, we identify these two strong absorption peaks in Cu$_2$O as transition involving the spin Zeeman split 2p$_{0}^{S_z=+1}$ and 2p$_{0}^{S_z=-1}$ levels from the orthoexciton state. Here, the $S_z=\pm1$ in the superscript indicate the projection of the total spin of the orthoexciton states along the field direction. This can be understood as follows; (i) the yellow exciton series in Cu$_2$O originate from an electric quadrupole transition, rather than a dipole allowed transition\,\cite{elliott1961Symmetry,artyukhin2018magneto}. In this case, only the orthoexciton states are optically active (bright), while the paraexciton states are forbidden (dark)\,\cite{artyukhin2018magneto}. The spin Zeeman nature of the splitting of the 2p$_{0}^{S_z=+1}$ and 2p$_{0}^{S_z=-1}$ orthoexciton states is confirmed by the $\sigma^+$ and $\sigma^-$ polarization resolved absorption. (ii) the splitting of the absorption peaks is close to the spin Zeeman energy in Cu$_2$O ($\simeq 2\mu_B B$ due to the absence of spin-orbit coupling in the conduction band and a hole g-factor which is close to zero\cite{artyukhin2018magneto}), but much smaller than the orbital Zeeman effect in Cu$_2$O (the \emph{exactly} $\hbar \omega_c$ splitting of the 2p$_{\pm 1}$ states is imposed by time reversal symmetry). (iii) two additional weak absorptions are observed as a shoulder-like structure above and below the 2p$_{0}^{S_z=\pm 1}$ peaks, suggesting 2p$_{+1}$ and 2p$_{-1}$ as the origin of these features. Therefore, the two pronounced absorption peaks have to be assigned to the 2p$_0^{S_z=\pm 1}$ excitonic states. Equally, the excited 3p transitions have to be assigned to the 3p$_0^{S_z=\pm 1}$ excitonic states. A detailed discussion of the assignment of the quantum numbers is provided in SI Sec.\,I. 

\begin{table}[b!]
  \caption{Summary of parameters used to calculate the hydrogenic transitions of Cu$_2$O. The exciton binding energy $Ry^*$ and the effective $g$-factor are in reasonable agreement with those obtained from DC magnetic field measurements\,\cite{artyukhin2018magneto}.}
  \label{Tbl_1}
  \setlength{\tabcolsep}{3mm}{
  \begin{tabular}{cccccc}
    \hline   \hline
    $E_g$ (eV) & $\mu^*$ ($m_e$) &  $R_y^*$ (meV) & $g$\\
    \hline
    2.1756 & $0.415\pm 0.01$ & $103 \pm 1$  & $2.0 \pm 0.1$ \\
    \hline \hline
  \end{tabular}}
\end{table}

Having assigned the quantum numbers of the absorptions, we use numerical solutions of the hydrogen model in a magnetic field to fit the data\,\cite{makado1986energy,Duncan2024Analytical}. The Makado and McGill calculations considers only the orbital Zeeman effect. In order to fit the data of Cu$_2$O we have included the spin Zeeman contribution. The orange and violet lines in Fig.\,\ref{Fig:energy2p3p} indicate the fit of 2p$_{0}^{S_z=+1}$ and 2p$_{0}^{S_z=-1}$ levels from the hydrogen model\,\cite{makado1986energy}. The fitting parameters are summarized in Table\,\ref{Tbl_1}. The exciton binding energy $Ry^*$ and the effective $g$-factor of Cu$_2$O used in the hydrogen model are in reasonable agreement with the DC field measurements\,\cite{artyukhin2018magneto}. Note that the value of the exciton binding energy $Ry^*$ is imposed by the measured zero field 2p-3p splitting which has to follow the $1/n^2$ Rydberg series. We cannot exclude that the exact value of the exciton binding energy is sample dependent which would explain the small reported variations (5\%).  The reduced exciton effective mass of $\mu^* = 0.415 \pm 0.01\,m_e$ is close to the high end of the range of values reported in the literature ($0.35-0.41\,m_e$)\,\cite{zhilich1969magnetoabsorption,artyukhin2018magneto,hodby1976cyclotron,halpern1967energy}. The hydrogen model fits very well to the four levels in Fig.\,\ref{Fig:energy2p3p} up to 300\,T. Above 300\,T the energy of the 2p$_0^{S_z=+1}$ state increases more slowly with magnetic field than the prediction of the hydrogen model, suggesting, in a first approximation, an increase of the reduced effective mass.


\begin{figure*}[t!]
  \centering
   \includegraphics[width= 0.5\linewidth]{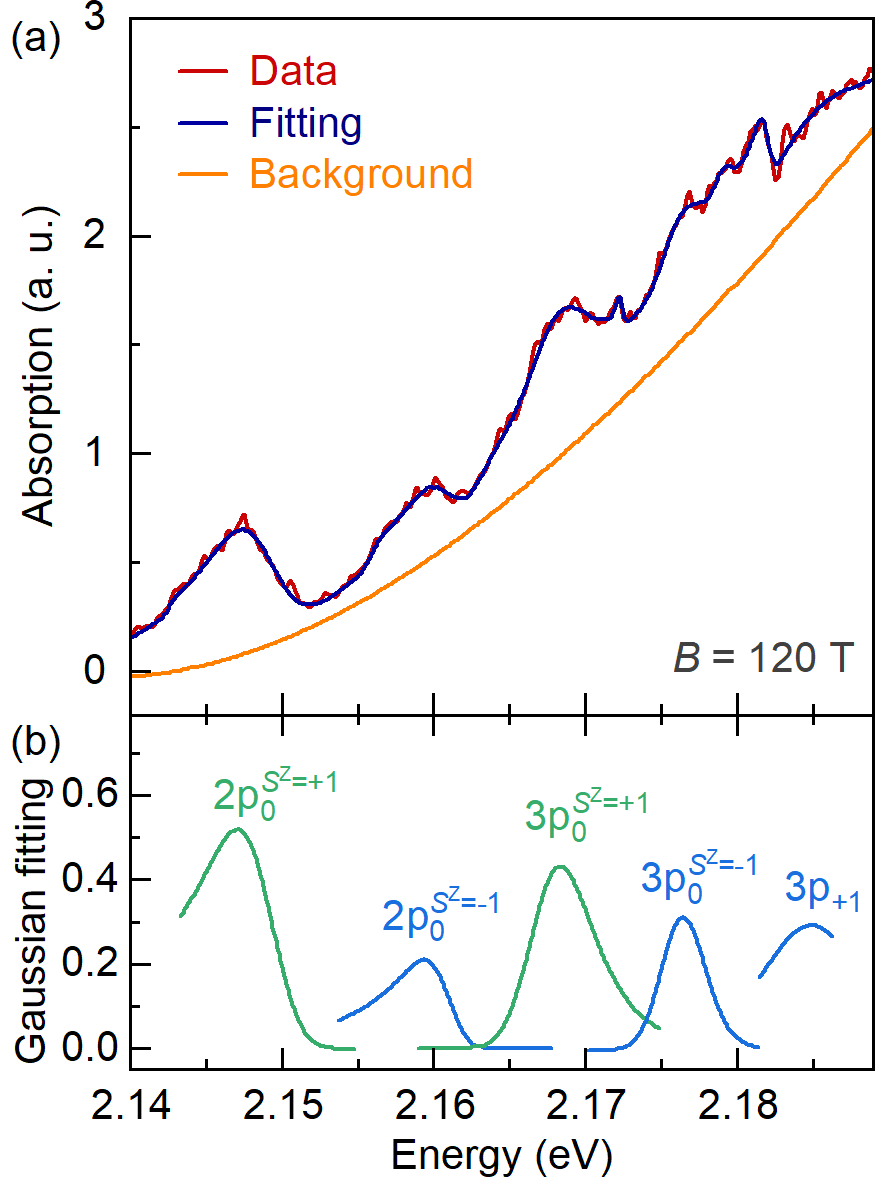}
  \caption{(a) Fit to the absorption spectrum of Cu$_2$O measured at $B=120$\,T and $T=45$\,K using multiple Gaussian functions. The red line is the measured absorption spectra, and the blue line is the superimposed fitting line. The orange line indicates the subtracted background used when making the Gaussian fits. (b) The multiple Gaussian function used to fit the experimental data. 
  }
  \label{Fig:Fit4FWHM}
\end{figure*}

\begin{figure*}[t!]
  \centering
   \includegraphics[width= 1\linewidth]{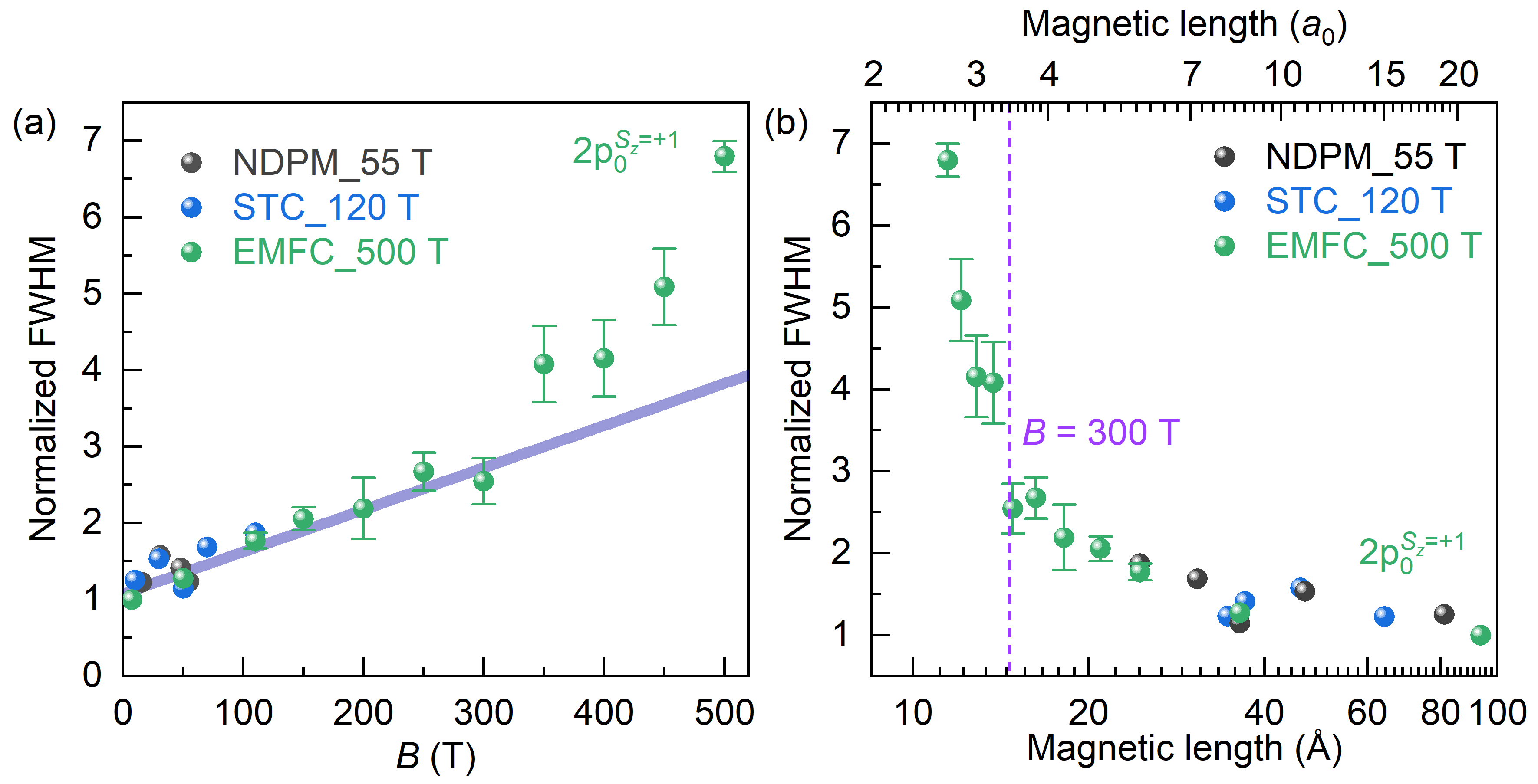}
  \caption{
  (a) FWHM of 2p$_0^{S_z=+1}$ excitonic absorption as a function of magnetic field normalized by the zero-field FWHM ($\simeq 2.9$\,meV). (b) Normalised FWHM of 2p$_0^{S_z=+1}$ excitonic absorption as a function of magnetic length $\ell_B = \sqrt{\hbar/eB}$.
  }
  \label{Fig:2p0-FWHM}
\end{figure*}

Having established that the energy shift of the excitonic states is generally well described by the hydrogen model, we now turn to investigate the evolution of the full width half maxima (FWHM) of the excitonic states. To do so, we fit the 2p and 3p excitonic absorption spectra using
asymmetrical Gaussian functions. An example is shown in Fig.\,\ref{Fig:Fit4FWHM}\,(a), the red line is the magneto-absorption spectrum measured at 120\,T, the orange line is the subtracted background, and the blue line is the fitting result. The Gaussian functions corresponding to each excitonic state are shown in Fig.\,\ref{Fig:Fit4FWHM}(b). We have performed this type of multi-peak fitting on all the magneto-absorption spectra up to 500\,T, and summarize the extracted FWHM as a function of magnetic field in Fig.\,\ref{Fig:2p0-FWHM} (a). 

Here we focus on the FWHM of the 2p$_{0}^{S_z=+1}$ transition which we can follow all the way up to 500\,T. The 2p$_{0}^{S_z=+1}$ level is a perfect candidate to investigate the influence of magnetic field on the excitonic line width, because it is well separated from other excitonic transitions (except close to zero magnetic field).  As seen in Fig.\,\ref{Fig:2p0-FWHM}\,(a), the FWHM of 2p$_{0}^{S_z=+1}$ shows a steady linear increase with the magnetic field up to 300\,T (see blue line). A linear dependence of the FWHM was also observed for the 1s excitonic state in III-V semiconductors\,\cite{polimeni2002linewidth}. However, at $B \geq 300$\,T, there is a departure from this linear behavior with a marked upturn in the FWHM of 2p$_{0}^{S_z=+1}$ transition which shows significant additional broadening. Simultaneously, the energy evolution of 2p$_{0}^{S_z=+1}$ deviates from the hydrogen model over the same magnetic field range, as can be seen in Fig.\,\ref{Fig:energy2p3p}.

In a magnetic field, the carriers undergo cyclotron motion with a radius which is characterized by the magnetic length $\ell_B = \sqrt{\hbar/eB}$\,\cite{miura2007physics}. Fig.\,\ref{Fig:2p0-FWHM}\,(b) shows the evolution of the FWHM as a function of the magnetic length plotted both in Angstroms (bottom axis) and as a multiple of the lattice constant (top axis). At a relatively low field, the magnetic length is much larger than the lattice constant and the electrons and holes are subjected to the average potential of the crystal (effective mass approximation). However, as the magnetic field increases, the magnetic length decreases. When the magnetic length is comparable to the crystal lattice constant, the effective mass approximation is expected to break down, and the electrons and holes start to feel the electrostatic potential of the individual atoms. In this case, the behavior of the carriers follows more closely the description of tight binding theory. According to the literature, Cu$_2$O crystallizes in a cubic structure with a lattice constant of $a_0=4.27$\,\AA\,\cite{Ruiz1997Electronic}. The abrupt broadening of the line width occurs around $B \simeq\,$300\,T, corresponding to a magnetic length of $\ell_B \simeq 14.8$\,\AA, which is only 3.5 times the lattice constant. It seems reasonable to conclude that, at least qualitatively, the sudden energy broadening at $B$\,=\,300\,T is due to the significant increase of the degeneracy of the energy levels that originate from the breakdown of the effective mass theory, namely, the Harper broadening\,\cite{harper1955single}. This hypothesis is supported by the observed deviation of the energy of the transition above 300\,T from the predicted variation of the hydrogen model, which also suggests a break down of the effective mass approximation at ultrahigh magnetic fields. 

The Harper broadening observed in a natural crystal suggests a significant competition between the periodic crystal lattice potential and the cyclotron energy. In higher magnetic fields, the effective mass approximation breaks down and even the chemical bonding between atoms in matter could be altered through the modification of the wave functions \cite{LangeScience2012}. The significant broadening observed at near 300\,T corresponding to $\alpha = \phi /\phi_0 \sim 0.013$, where $\phi$ is the magnetic flux and $\phi_0 = h/e$ is the flux quantum, provides a benchmark for the ultrahigh magnetic field control of the physics in a crystal. For example, $\alpha$ controls the fractal structure of Hofstadter's butterfly \cite{HofstadterPRB1976}. 



\section{Conclusions}

In conclusion, we have performed magneto-absorption measurements on the famous yellow-exciton series in Cu$_2$O using various pulsed field generation techniques, producing magnetic fields of up to 500\,T. The ultra-high field absorption measurement allows us to correctly assign the quantum numbers of lowest energy absorption peaks identifying transitions to the spin Zeeman split of the 2p$_{0}$ state. The ultrahigh field data provides a value for the reduced mass of the exciton $\mu^* = 0.415 \pm 0.01 m_e$. Intriguingly, the FWHM of 2p$_{0}^{S_z=+1}$ transition exhibit a sudden increase at magnetic fields above 300\,T, providing possible evidence for Harper broadening - an indication of the break-down of the effective mass approximation when the magnetic length becomes comparable to the lattice constant of the crystal.

\section{Method}
\subsection{Sample}

The Cu$_2$O single crystal was purchased from Crystal Base Co. Ltd with a crystal surface of (0 0 1). To obtain a measurable transmission spectrum, the Cu$_2$O single crystal was mechanically polished down to 48\,$\upmu$m, and glued to a quartz disk by transparent epoxy (Loctite Stycast 1266J). The sample was cut down to 1.8$\times$1.8\,mm to allow the same crystal to be used with all the available techniques to generate pulsed magnetic field.

\subsection{Magneto-absorption in non-destructive pulse magnet}
For the magneto-absorption measurement with the magnetic fields up to 60\,T, a non-destructive pulsed magnet was used, with a typical pulse duration of 36\,ms. The sample was mounted in a liquid helium cryostat. A broadband white light from a halogen lamp was used as the excitation; the transmitted light from the sample was guided by an 800\,$\upmu$m multi-mode fiber to the spectrometer coupled with a CCD camera. The typical integration time 0.5\,ms, which ensured each spectrum was acquired at essentially constant magnetic field. The $\sigma^+$ and $\sigma^-$ polarization was resolved by inserting a circular polarizer between the sample and the white light source. The magnetic field direction was reversed to select between $\sigma^\pm$ polarization.

\subsection{Magneto-absorption in Single-turn coil system}
For the magneto-absorption measurement with the magnetic field up to 200\,T, the magnetic field pulses with a typical duration of 10\,$\upmu$s were generated by a single-turn coil system with a bore diameter of 10\,mm. A specially designed helium-flow type cryostat was used to cool the sample down to 10\,K. A xenon arc-flash lamp was used as the light source. The light was guided and collected by 800\,$\upmu$m multi-mode fibers. The time dependence of the transmitted light was measured by a high-speed streak camera coupled with a polychromator.  The $\sigma^+$ and $\sigma^-$ polarization was resolved by using left-hand and right-hand circular polarizers.

\subsection{Magneto-absorption in electromagnetic flux compression}
For the magneto-absorption measurement in ultra-high magnetic fields of up to 520\,T, the electromagnetic flux compression (EMFC) method was used to generate magnetic field pulses with a typical duration of 50\,$\upmu$s\,\cite{nakamura2018record}. The optical setups that coupled to EMFC equipment are essentially the same as the one coupled to the single-turn coil system.


\begin{acknowledgments}

This study has been partially supported through the EUR grant NanoX no. ANR-17-EURE-0009 in the framework of the ``Programme des Investissements d’Avenir''. Z.Y. was supported by research grant from Research Foundation for Opto-Science and Technology and JSPS KAKENHI Grant Number 25K17324
(Grant-in-Aid for Early-Career Scientists). Y.I. and Y.H.M. were supported by the JSPS KAKENHI, Grant-in-Aid for Transformative Research Areas (A) Nos.23H04859 and 23H04860. A.M. was funded by JSPS KAKENHI, Fund for the Promotion of Joint International Research (Home-Returning Researcher Development Research) No. 22K21359.

\end{acknowledgments}


\bibliography{JoeyBibOne4All.bib}

\end{document}




\title{Supplemental Material for: Possible evidence for Harper broadening in the yellow series exciton of \texorpdfstring{Cu$_2$O}{Cu2O} at ultrahigh magnetic field}

\author{Zhuo Yang}
\affiliation{Institute for Solid State Physics, The University of Tokyo, Kashiwa, Chiba, 277-8581, Japan}

\author{Jinbo Wang}
\affiliation{Institute for Solid State Physics, The University of Tokyo, Kashiwa, Chiba, 277-8581, Japan}

\author{Yuto Ishii}
\affiliation{Institute for Solid State Physics, The University of Tokyo, Kashiwa, Chiba, 277-8581, Japan}

\author{Duncan K. Maude}
\affiliation{Laboratoire National des Champs Magn\'etiques Intenses, CNRS-UGA-UPS-INSA, 143 avenue de Rangueil, 31400 Toulouse, France}

\author{Atsuhiko Miyata}
\affiliation{Institute for Solid State Physics, The University of Tokyo, Kashiwa, Chiba, 277-8581, Japan}

\author{Yasuhiro H. Matsuda}
\email{ymatsuda@issp.u-tokyo.ac.jp}
\affiliation{Institute for Solid State Physics, The University of Tokyo, Kashiwa, Chiba, 277-8581, Japan}

\date{\today}

\newcommand{\Joey}[1]{{\color[RGB]{26,111,223}{#1}}}
\newcommand{\Joeyc}[1]{{\color[RGB]{203,5,5}{[Joey: {#1}\,]}}}
\newcommand{\Joeyx}[1]{{\color[RGB]{55,173,107}{\st{#1}}}}

\newcommand{\Yhmc}[1]{{\color[RGB]{184,0,184}{[Matsuda: {#1}\,]}}}

\clearpage

\maketitle

\clearpage

\section{A\MakeLowercase{ssigning the quantum numbers of the {2p} excitonic absorption features}}
\label{SI:Sec:Quantum number}

\begin{figure*}[h!]
  \renewcommand \thefigure {S\arabic{figure}}
  \centering
   \includegraphics[width= 0.95\linewidth]{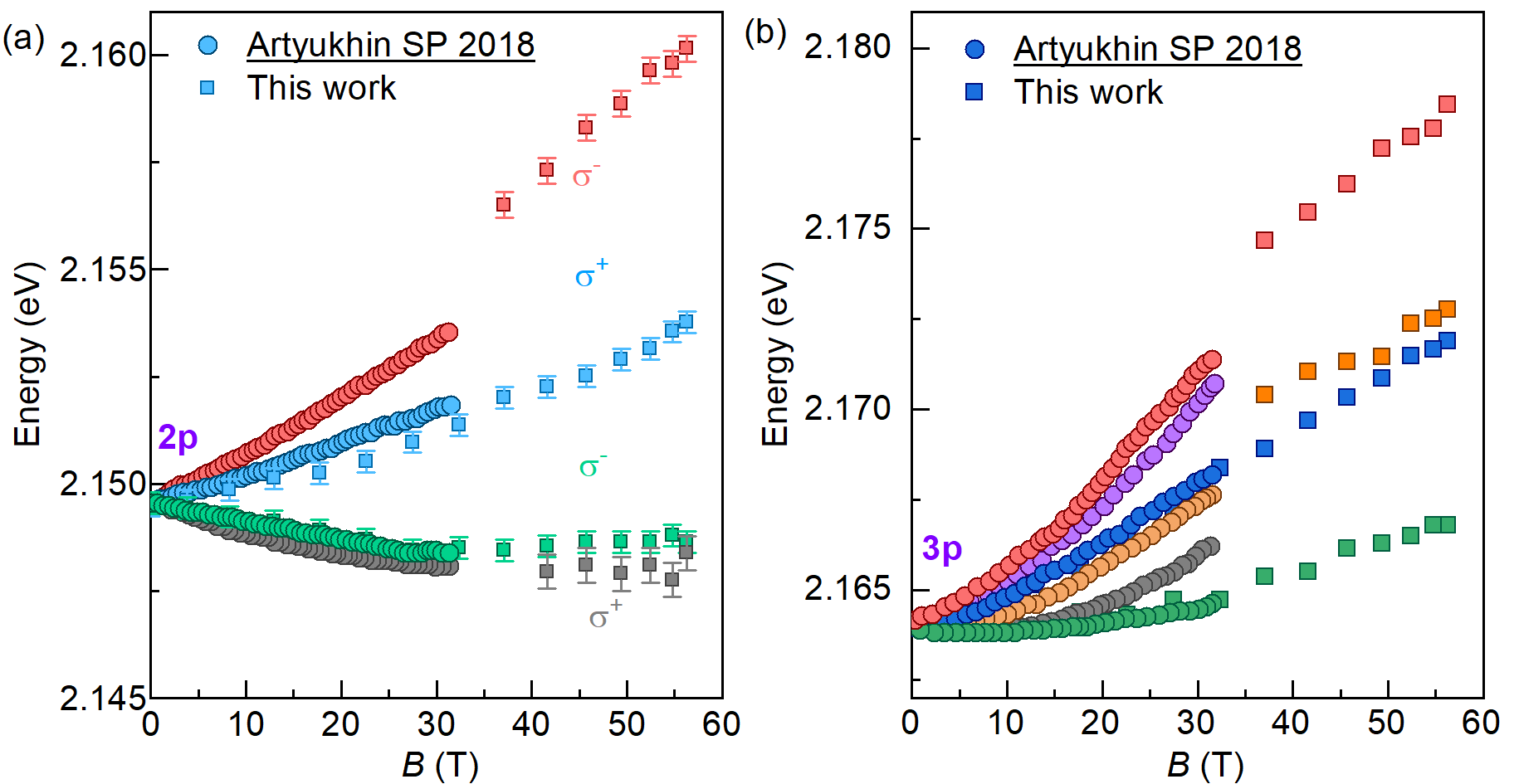}
  \caption{
Circular symbols are the digitized data points of 2p excitonic absorption energy (a) and 3p excitonic absorption energy (b) at \textit{T}\,=\,1.2\,K as a function of magnetic field up to 31\,T from ref\,\cite{artyukhin2018magneto}, while the square symbols are the excitonic absorption energy measured at 1.5\,K and field of up to 57\,T from this work. 
}
  \label{SI:Fig:DatapointsCom}
\end{figure*}

Under magnetic field, 2p excitonic level splits into 2p$_{-1}$, 2p$_{0}$ and 2p$_{+1}$ levels due to the orbital Zeeman effect; here, the $0$ and $\pm 1$ in the subscript represent the magnetic quantum number $m$. Due to the spin Zeeman effect, these states are further classified as either an orthoexciton (triplet state with a total spin of $S=1$ and projection along the $z$ axis $S_z=0,\pm 1$) or a paraexciton (singlet state with a total spin of $S=0$ and $S_z=0$). Under magnetic field ($B\|z$), the orthoexciton splits into three levels with $S_z = 0,\pm 1$, while the degeneracy of the singlet paraexciton state with $S_z=0$ will be lifted if the electron and hole $g$-factors are not identical. In solid state physics the singlet/triplet states are referred to as exciton fine-structure, which for s-states can exhibit a zero magnetic field splitting when the exchange interaction is taken into account. However, for the p-orbitals there is no exchange contribution so the zero field splitting can be safely neglected \cite{artyukhin2018magneto}. 

Not all the levels are expected to be optically active when the selection rules are taken into consideration. The interpretation of the origin of the 2p energy levels of Cu$_2$O in magnetic field and assignment of the quantum numbers are not fully consistent in the literature\,\cite{kobayashi1989yellow,artyukhin2018magneto,zielinska2019magneto}. In this work, we combined the circular polarization, selection rules and hydrogen model in a magnetic field to assign the quantum number of the split 2p levels in Cu$_2$O.


The yellow exciton series in Cu$_2$O originates from the optical transitions from valence 3$d$ bands to conduction 4$s$ bands\,\cite{elliott1961Symmetry}. Having the same parity, the dipole transitions are forbidden and the yellow exciton series originate from an electric quadrupole transition. For this reason, only p-states are optically allowed (no transitions involving s-states are observed). When the symmetry of the yellow excitons is taken into account, theoretical considerations indicate that paraexciton (singlet) state is optically inactive, while the orthoexciton (triplet) state is optically allowed\,\cite{artyukhin2018magneto}. In this case, only nine levels of the 2p states can be observed in the magneto-optical measurements, namely 2p$_{-1}^{S_z=0,\pm1}$, 2p$_{0}^{S_z=0,\pm1}$ and 2p$_{+1}^{S_z=0,\pm1}$.

\begin{figure*}[h!]
  \renewcommand \thefigure {S\arabic{figure}}
  \centering
   \includegraphics[width= 0.5\linewidth]{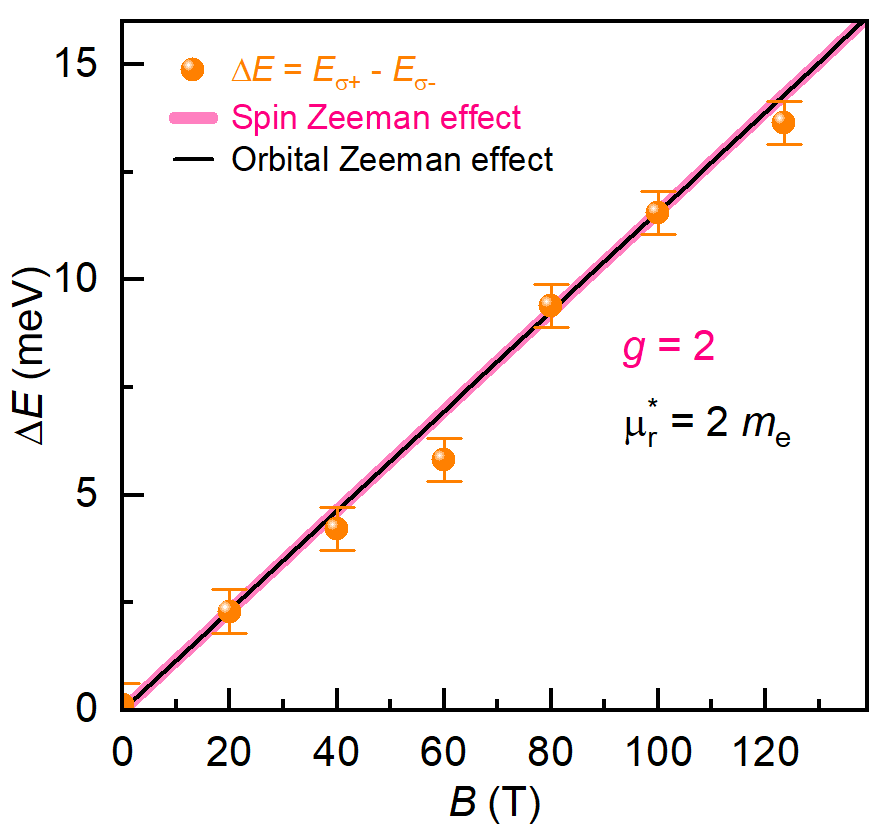}
  \caption{
 Splitting energies $\Delta E$ for the two main peaks in $\sigma^+$ and $\sigma^-$ polarization as a function of magnetic field up to 120\,T ($\Delta E$ between green and blue arrows in Fig.\,\ref{SI:Fig:NDPM_Spe}(a)). The pink and black lines represent the fit of spin and orbital Zeeman term, respectively.
}
  \label{SI:Fig:SpinOrbital}
\end{figure*}

The two main 2p absorption peaks (marked as blue and green arrows in Fig.\,\ref{Fig:FieldSpectra}(a)) can be separately distinguished using $\sigma^+$ and $\sigma^-$ polarized light. The angular momentum of the $\sigma^\pm$ photon can be used to either change the orbital angular momentum or flip an electron spin. Therefore, there are two possible origins for these main peaks, either 2p$_{0}^{S_z=\pm1}$ (spin flip transition) or 2p$_{\pm1}^{S_z=0}$ (orbital transition equivalent to a change in Landau level index for free carriers). Fig.\,\ref{SI:Fig:SpinOrbital} shows the splitting energies $\Delta E = E_{\sigma^+} - E_{\sigma^-}$ of the two strong absorption peaks as the function of magnetic field up to 120\,T. Assuming the two main peaks originate from an orbital Zeeman effect, a reduced effective mass value of $\mu_r^*$\,=\,2\,$m_\text{e}$ is obtained, which is five times higher than the accepted literature value (0.35\,-\,0.41\,$m_\mathrm{e}$)\,\cite{artyukhin2018magneto,zhilich1969magnetoabsorption,hodby1976cyclotron,halpern1967energy}. On the other hand, assuming these peaks originates from spin flip transition, we obtain an effective $g$-factor of $\simeq 2$, in consistent with the literature reports\,\cite{artyukhin2018magneto}. It is, therefore, reasonable to conclude that the two main peaks originates from the 2p$_{0}^{S_z=\pm1}$ states, and the shoulder-like features originates from 2p$_{\pm1}$ states.

\newpage

\section{E\MakeLowercase{nergies versus field diagram for 2p and 3p states}}

\begin{figure*}[h!]
  \renewcommand \thefigure {S\arabic{figure}}
  \centering
   \includegraphics[width= 0.75\linewidth]{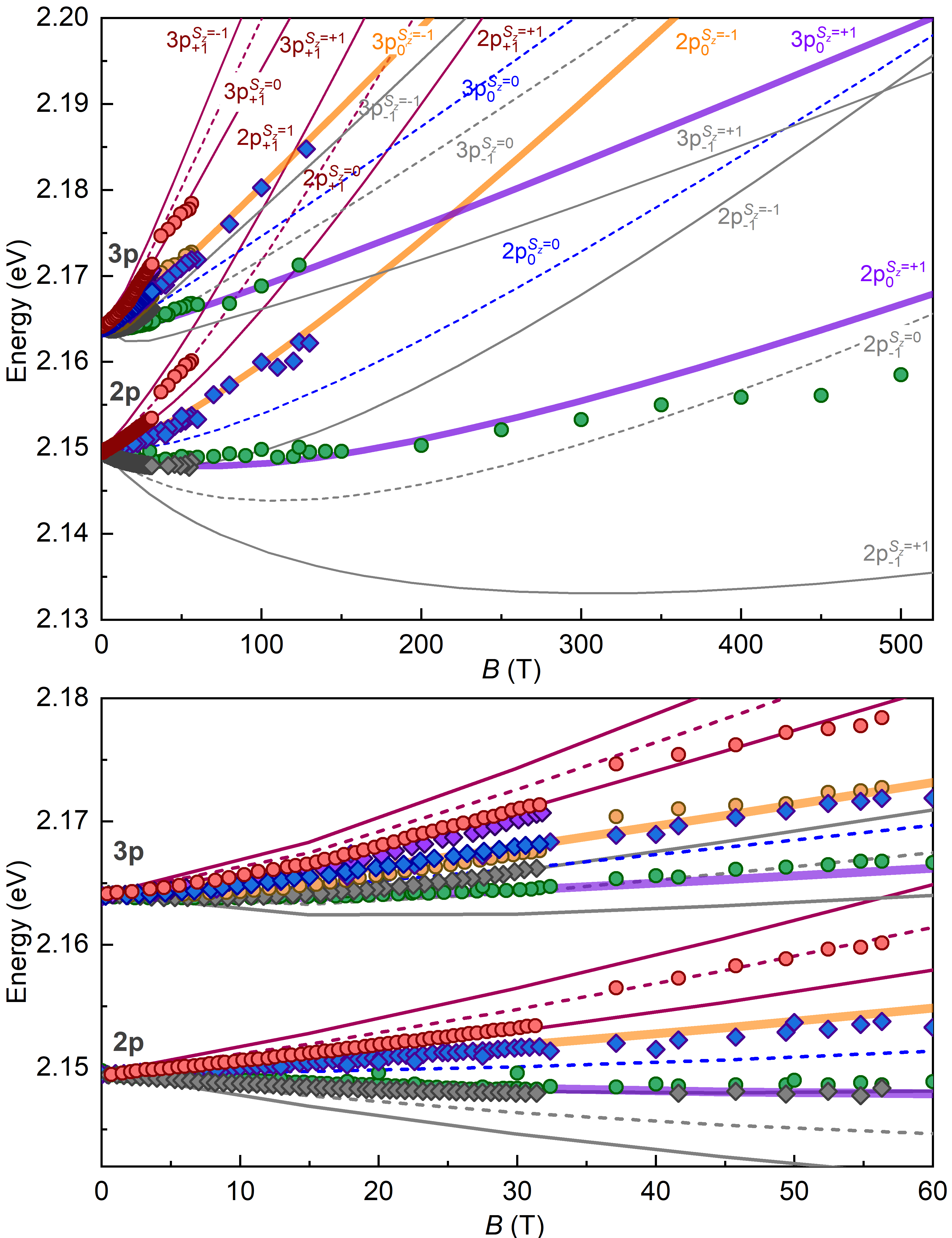}
  \caption{
 (a) Energies for 2p and 3p excitonic absorption at \textit{T}\,=\,45\,K as a function of magnetic field up to 500\,T. The circular and square symbols represent the absorption energy obtained from $\sigma^-$ and $\sigma^+$ polarization, respectively. The solid and dashed lines represent the fit from the hydrogen model in magnetic field. (b) Enlarged energies versus magnetic field diagram highlighting the fit in the relatively low field regime. The data points includes the results from NDPM, STC and EMFC techniques and also the literature\,\cite{artyukhin2018magneto}.
}
  \label{SI:Fig:Compare30to500T}
\end{figure*}

In addition, we used the hydrogen model in the magnetic field\,\cite{makado1986energy, Duncan2024Analytical} to globally fit all the split levels from 2p and 3p states, as shown in Fig.\,\ref{SI:Fig:Compare30to500T}. The fit and the experimental data show excellent agreement (see also the insert of Fig.\,\ref{SI:Fig:Compare30to500T}). The exciton binding energy, effective mass and effective $g$-factor used in the hydrogen model are $Ry^*$\,=\,103 $\pm$ 1\,meV, $\mu_\mathrm{r}^*$\,=\,0.415 $\pm$ 0.01$m_\mathrm{e}$ and $g$\,=\,2 $\pm$ 0.1, respectively, in reasonable agreement with the literature reports\,\cite{kazimierczuk2014giant,artyukhin2018magneto}, which justify our hypothesis. 

As mentioned in the previous section, all the data points can be resolved from the $\sigma^+$ and $\sigma^-$ polarizations, therefore, the two shoulder-like features are expected to be 2p$_{\pm1}^{S_z=0}$. However, it is clearly seen from the Fig.\,\ref{SI:Fig:Compare30to500T}\,(b) that the energy-field dependence of these shoulder peaks are more closely fitted by the 2p$_{+1}^{S_z=-1}$ and p$_{-1}^{S_z=+1}$ states. The fits of hydrogen model indeed imply that, regardless of the magnetic quantum number, the $\sigma^+$ and $\sigma^-$ polarized light only couples strongly with the spin flip transitions. 


\clearpage

\bibliography{JoeyBibOne4All.bib}